\documentclass[12pt]{article}

\usepackage{overcite}
\usepackage{graphicx}
\usepackage{dcolumn}

\usepackage[latin1]{inputenc}
\usepackage[T1]{fontenc}
\usepackage{ae,aecompl}

\begin{document}

\newcommand{\rar}{$\rightarrow$}
\newcommand{\lrar}{$\leftrightarrow$}

\newcommand{\beq}{\begin{equation}}
\newcommand{\eeq}{\end{equation}}
\newcommand{\bea}{\begin{eqnarray}}
\newcommand{\eea}{\end{eqnarray}}
\newcommand{\Req}[1]{Eq. (\ref{E#1})}
\newcommand{\req}[1]{(\ref{E#1})}
\newcommand{\degree}{$^{\rm\circ} $}
\newcommand{\pcite}{\protect\cite}
\newcommand{\pref}{\protect\ref}
\newcommand{\Rfg}[1]{Figure \ref{F#1}}
\newcommand{\rfg}[1]{\ref{F#1}}
\newcommand{\Rtb}[1]{Table \ref{T#1}}
\newcommand{\rtb}[1]{\ref{T#1}}

\vskip 0.5cm
\title{\large\bf Comments on "Remeasuring the Double Helix"}
\vskip 2.0cm
\author{Alexey K. Mazur\\
\small CNRS UPR9080, Institut de Biologie Physico-Chimique\\
\small 13, rue Pierre et Marie Curie, Paris,75005, France
}
\date{}

\maketitle
 

\begin{abstract}
Mathew-Fenn et al. (Science {\bf 2008}, 322, 446-9) measured
end-to-end distances of short DNA and concluded that stretching
fluctuations in several consecutive turns of the double helix should
be strongly correlated. I argue that this conclusion is based on
incorrect assumptions, notably, on a simplistic treatment of the
excluded volume effect of reporter labels. Contrary to the author's
claim, their conclusion is not supported by other data.
\end{abstract}

Long DNA double helices behave as continuous elastic rods placed in a
heat bath. \cite{Cantor:80} The average end-to-end distance
$R$ and its fluctuations depend upon the temperature, the molecular
elasticity and the DNA length $L$. The fluctuations are commonly
characterized by one of the two ascending functions, namely, the
standard deviation, $\sigma(L)$, or the variance, $\sigma^2(L)$. The
growth of $\sigma(L)$ is mainly due to bending. The twisting
contribution to $\sigma(L)$ is oscillating and it does not accumulate
with $L$. However, in fragments much shorter than the bending
persistence length ($l_b$=50 nm) the relative contribution of stretching may
become appreciable. If the local stretching fluctuations in a straight
double helix are statistically independent (as in the elastic rod),
the variance $\sigma^2(L)$ should be a linear function. A convex shape
of $\sigma^2(L)$ indicates that local fluctuations are correlated.

The striking conclusion of Mathew-Fenn et al. \cite{Mathew-Fenn:08b}
is based upon a convex (nearly square) measured profile of
$\sigma^2(L)$.  This observation is corroborated by similar
dependences obtained from the published data for two alternative
methods (trsmFRET and DERR). One more method can be added to this
list, namely, the atomic force microscopy (AFM). In AFM, the DNA
contour length is measured directly by the Gaussian fitting to the
distribution produced by a large number of DNA images. Recent studies
demonstrated that, in optimal conditions, good agreement with the
canonical B-DNA length is reached,
\cite{Rivetti:01,Sanchez-Sevilla:02,Podesta:04} with the relative
standard deviation $\sigma(L)/L$ nearly constant in the range of DNA
lengths from 300 to 4000 bp, corresponding to a square growth of the
variance.

A closer look reveals, however, that in all these cases a convex shape
of $\sigma^2(L)$ cannot be due to DNA and rather should be attributed
to the method. This becomes clear from comparison of the fluctuation
amplitudes.  The relative standard deviation $\sigma(L)/L$ is
about 7 \% for the author's new method, but 17 \% for the
trsmFRET and DERR, and 3 \% for AFM. Three different values
cannot all refer to the true DNA fluctuations. Rather the smallest of
them obtained with AFM establishes the upper limit because experimental
errors augment the amplitudes. The contribution of DNA dynamics in the
AFM variance is unknown and probably small, but we can be sure that in
the other three methods the variances are dominated by factors other
than DNA itself. If the postulated correlations occur only with $L<35$
bp the reported $\sigma^2(L)$ dependence should be continued linearly
on $L>35$bp; the $\sigma(L)/L$ decreases, and the AFM level is reached
with $L\approx300$bp.  This is the lower boundary of the available AFM
data (to my knowledge), but the linear $\sigma(L)$ profile in AFM is
stable over hundreds of bp and there is no reason to suspect that it
stops at 300 bp. With some reservations concerning the accuracy of the
literature data, we have to conclude that the above discrepancies
argue against the author's interpretation of the convex profile of
$\sigma^2(L)$.

It is more reasonable to suppose that the $\sigma^2(L)$ profiles in
all above examples are dominated by other experimental factors. In the
present case, the authors erroneously assumed that the bulky labels
attached to DNA make a constant contribution to $\sigma^2$.  This
would be true if the labels were freely rotating around the points of
attachment. In reality, these are bulky objects that, due to the
excluded volume effect, on average protrude from DNA and continue it.
Their motion effectively increases the bending contribution to the
measured variance of $R(L)$. Quantitatively this effect is evaluated in
\Rfg{alvsL} where the data of Ref.  \citeonline{Mathew-Fenn:08b} are
compared with BD simulations of discrete worm-like chain (WLC) models
of the corresponding labeled DNA fragments. The BD method correctly
samples from the canonical ensembles with correct thermodynamic
averages \cite{Mzjpc:08}. The gold clusters were modeled as two beads
attached to the opposite DNA ends on 1.5 nm strings, which
approximately corresponds to their true size and provides satisfactory
agreement with experimental $R(L)$ regardless of the fitting
parameters (see the upper panel of \Rfg{alvsL}). The stretching
modulus for was constant for the whole system ($\approx$1000pN). The
bending persistence length of DNA $l_b$ and the average deviation
angle of the end labels $\bar\theta$ were varied separately.

\begin{figure}[ht]
\centerline{\includegraphics[width=8cm]{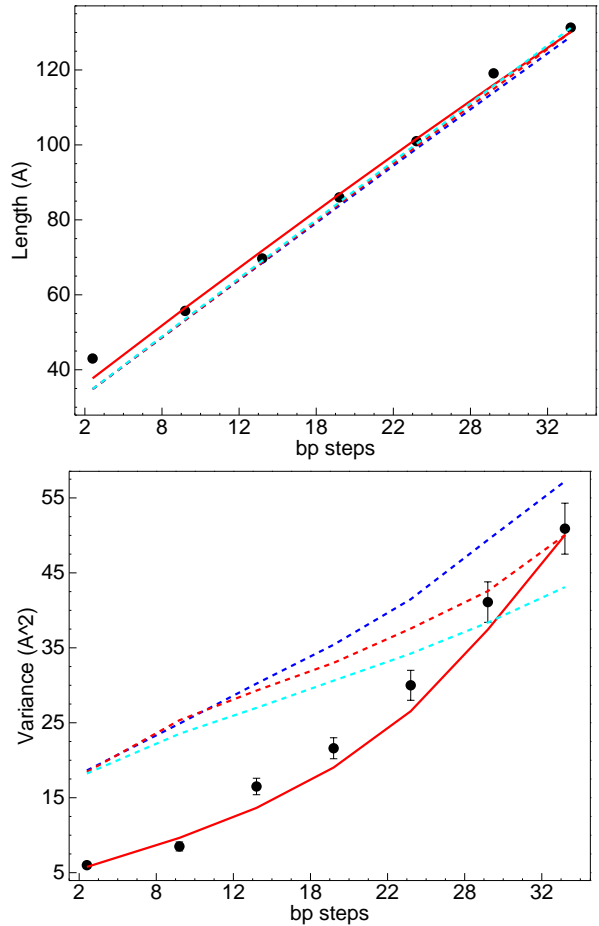}}
\caption{\footnotesize\label{FalvsL}
Comparison of experimental data reported in the supplementary
material of Ref. \citeonline{Mathew-Fenn:08b} with BD simulations of
discrete WLC models described in the text. Black circles with error
bars show the experimental data.
}\end{figure}

The dashed red lines in \Rfg{alvsL} display the results for the
standard DNA persistence length $l_b$=50 nm, and $\bar\theta$ adjusted
to fit the experimental $\sigma^2(35\rm bp)$, which is achieved with
$\theta\approx46\pm20^\circ$. This corresponds to relatively free
motion, but the labels never look backward with respect to DNA
direction. The bending contribution to the variance for 35bp unlabeled
DNA is about 7 \AA$^2$. With the labels attached, it is greatly
increased due to the end fluctuations enhanced by the flexibility of
DNA.  This is seen from the difference between the blue, red, and cyan
traces corresponding to $l_b$=40, 50, and 60 nm, respectively, with
the same $\theta$. For very short DNA the three traces converge to the
value corresponding to the proper contribution of the attached
clusters. It is larger than in experiment because, with small $R(L)$,
the clusters interact directly. The cluster-cluster excluded volume
effect significantly restrains their motion, which can be readily
shown by introducing appropriate potentials (not shown).  \Rfg{alvsL}
checks how strong these restraints should be to fit the experimental
data. To this end, the $\bar\theta$ value was was fit to experiment
for L=4bp and the corresponding energy coefficients were linearly
interpolated for 4bp<L<35bp. The result is shown in \Rfg{alvsL} by the
solid red line. The corresponding $\bar\theta$ values gradually
increase from 30$^\circ$ to 46$^\circ$, which is not large.

As we see, these experimental data can be well accounted for with a
few physically reasonable assumptions. The most important of them
concerns the cluster-cluster interactions, however, it is supported by
the published distance distribution for these clusters joined by a
flexible hinge. \cite{Mathew-Fenn:08a} Notably, the contact distance
is $\ge$40 \AA, which corresponds to the diameter of a passivated
cluster plus two layers of hydration water. The same data
\cite{Mathew-Fenn:08a} suggest that the cluster-cluster interactions
are long-range and may well continue to 100 \AA. The fitting in
\Rfg{alvsL} demonstrates that $\sigma^2$ is very sensitive to the
amplitude of the end fluctuations. The clusters are attached to DNA by
flexible hinges, therefore, the actual value of $\bar\theta$ would be
affected by weak environmental factors, like electrostatic and
hydration end effects that may slightly evolve with the DNA length.

One can note in conclusion that, although the new molecular ruler
developed by the authors opens interesting new prospects in studying
biomacromolecules, neither their data nor the earlier literature
indicate the existence of long-range stretching fluctuations in double
helical DNA.

\centerline{\large\bf Appendix}

{\em The foregoing Technical Comments were considered for publication
in Science during four months and rejected based upon the
considerations given in the letter below. Regrettably, the Science
editors were strongly against posting this information and I had to
drop the referee's comments.}\\

\noindent
13 April 2009

\noindent
Dear Dr. Mazur:

Thank you for submitting your Technical Comment on the Science paper
by Mathew-Fenn et al.

We sent your comment and the author response, which is attached for
your information, out to two external referees for evaluation.
Unfortunately their reviews (appended below) are not positive enough
to support publication of your manuscript. As you'll see, although the
referees felt that your comment raised potentially valid technical
points, they did not find the development of those points sufficiently
clear and persuasive to drive the debate forward and provide an
enlightening discussion for the broad readership of Science.  

Although we recognize that you may be able to address some of the
specific criticisms in a revised manuscript, the overall nature of the
reviews is such that the manuscript would not be able to compete with
other Technical Comments under consideration. Notwithstanding this
disappointing outcome, we appreciate the chance to consider the
comment, and hope that you find the authors' response and the referee
comments helpful should you decide to prepare the manuscript for
submission to another journal. Thanks for your patience during this
long process, and thanks for your interest in Science.

Sincerely,

\noindent
Tara S. Marathe\\
Associate Online Editor, Science\\
tmarathe@aaas.org

\end{document}